# A Performance Study of Monitoring and Information Services for Distributed Systems

Xuehai Zhang[1], Jeffrey Freschl[2] and Jennifer M. Schopf[3]
hai@cs.uchicago.edu, jefflf@cats.ucsc.edu, jms@mcs.anl.gov

[1]Department of Computer Science, University of Chicago
[2]Department of Computer Science, University of California at Santa Cruz
[3]Mathematics and Computer Science Division, Argonne National Laboratory

## ABSTRACT

*Monitoring and information services form a key component of a distributed system, or Grid. A quantitative study of such services can aid in understanding the performance limitations, advise in the deployment of the systems, and help evaluate future development work. To this end, we study the performance of three monitoring and information services for distributed systems: the Globus Toolkit's Monitoring and Discovery Service (MDS), the European Data Grid Relational Grid Monitoring Architecture (R-GMA) and Hawkeye, part of the Condor project. We perform experiments to test their scalability with respect to number of users, number of resources and amount of data collected. Our study shows that each approach has different behaviors, often due to their different design goals. In the four sets of experiments we conducted to evaluate the performance of the service components under different circumstances, we found a strong advantage to caching or pre-fetching the data, as well as the need to have primary components at well connected sites due to high load seen by all systems.*

## 1. INTRODUCTION

Grid [7] platforms require and depend on monitoring and information services to support the discovery, monitoring and management of the distributed resources for a variety of tasks. For example, a user may want to determine the best platform to run an application on, a client program may want to collect a stream of data to help steer an application, or a system administrator may want to be notified when changes in system load or free disk space occur. Because of this, it is helpful to study the behavior of a monitoring and information services under different circumstances in order to understand any performance limitations in common settings.

To date, very few results have been published that quantitatively study the performance of the current monitoring and information services in distributed systems. Aloisio *et al.* [1] studied the capabilities and limitations of the Globus Toolkit's Monitoring and Discovery Service; however, their experiments were limited to simple tests on a GIIS only. Plale, Dinda and Laszewski [17] examined the advantages and disadvantages of building a Grid Information Service on a hierarchical representation and a relational or flat table representation. However, the comparison was based on the discussion of the different features of concepts and services used in these two representations. A recent study by Plale [16] benchmarked a synthetic workload formed by a set of basic operators (queries, updates, etc.) against a theoretical information service implemented using two different databases, MySQL and Xindice, and then evaluated the performance. Currently no information service uses either of these approaches though.

In this paper, we describe the scalability and performance of three monitoring and information services: the Globus Monitoring and Discovery Service (MDS) [2][14], the Relational-Grid Monitoring Architecture (R-GMA) [4] used in the European DataGrid [3], and Hawkeye [10], part of the Condor Project [13]. Each of these systems is in use in production or near-production grid testbeds. To facilitate the performance comparison among the three services, we map the functional components of one service to the counterparts of the others. Details of each are given in Section 2.

We designed a set of experiments to evaluate the effect of a large number of users, resources, and information sources for each service. We analyzed the performance limiting factors and their influence, and compared the performance behavior.

The analysis and comparison results, given in Section 3, also provide constructive suggestions for performance improvements. Section 4 gives the conclusions and briefly discusses future work.

## 2. MDS, R-GMA and Hawkeye

This section describes the background of the three monitoring and information services and the mapping of the functional components among them.

### 2.1 MDS

The Monitoring and Discovery Service (MDS) [2][14] is the Grid information service used in the Globus Toolkit [6]. It uses an extensible framework for managing static and dynamic information about the status of a computational Grid and all its components: networks, compute nodes, storage systems, instruments, and so on. MDS is built on top of the Lightweight Directory Access Protocol (LDAP) [15][21].

MDS is primarily used to address the resource selection problem, namely, how does a user identify the host or set of hosts on which to run an application? MDS is designed to provide a standard mechanism for publishing and discovering resource status and configuration information. It provides a uniform, flexible interface to data collected by lower-level information providers. It has a decentralized structure that allows it to scale, and it can handle static or dynamic data about resources, queues and the like. With MDS, one can also restrict access to data using GSI (Grid Security Infrastructure) credentials. MDS has a hierarchical structure (see Figure 1) that consists of three main components. A *Grid Index Information Service (GIIS)* provides an aggregate directory of lower level data. A *Grid Resource Information Service (GRIS)* runs on a resource and acts as a modular content gateway for a resource. *Information Providers (IPs)* interface from any data collection service and then talk to a GRIS. Each service registers with others using a soft-state protocol that allows dynamic cleaning of dead resources. Each level also has caching to minimize the transfer of un-stale data and lessen network overhead.

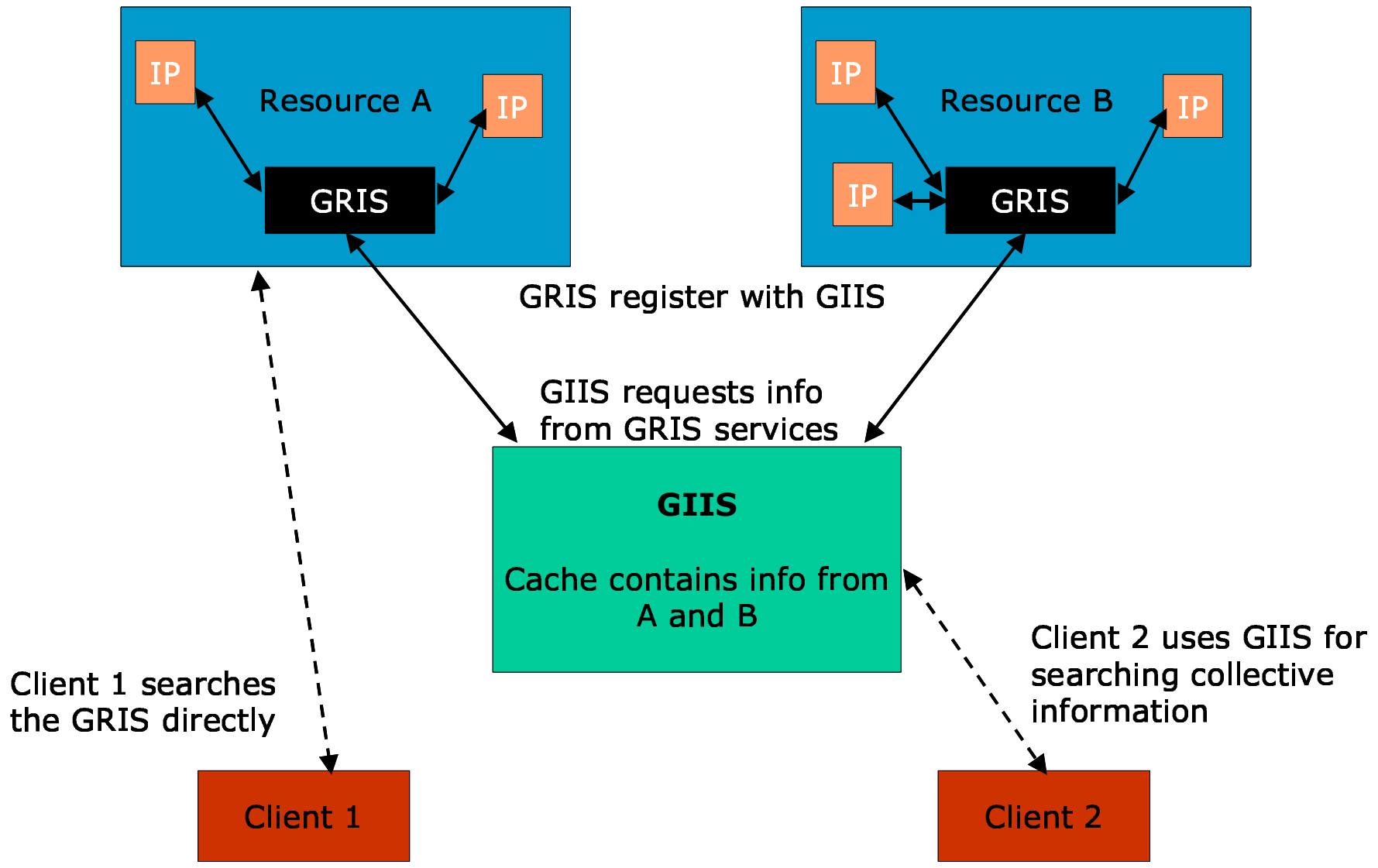

**Figure 1: The MDS architecture is a flexible hierarchy. There can be several levels of GIISs, and any GRIS or GIIS can register with another, making this approach modular and extensible.**

### 2.2 GMA and R-GMA

The Relational Grid Monitoring Architecture (R-GMA) [4] monitoring system is an implementation of the Grid Monitoring Architecture (GMA) [20] defined within the Global Grid Forum (GGF) [9]. It is based on Relational Database Management System (RDBMS) [5] and Java Servlet technologies [11]. Its main use is the notification of events—that is, a user can subscribe to a flow of data with specific properties directly from a data source. For example, a user can subscribe to a load-data data stream, and create a new Producer/Consumer pairing to allow notification when the load reaches some maximum or minimum.

GMA is an architecture for monitoring components that specifically addresses the characteristics of Grid platforms. The GMA consists of three components (shown in Figure 2): Consumers, Producers, and a Registry. Producers register themselves with the Registry and Consumers query the Registry to find out what types of information are available and to locate the corresponding Producers. Then the Consumer can contact a specific Producer directly. GMA as defined currently does not specify the protocols or the underlying data model to be used.

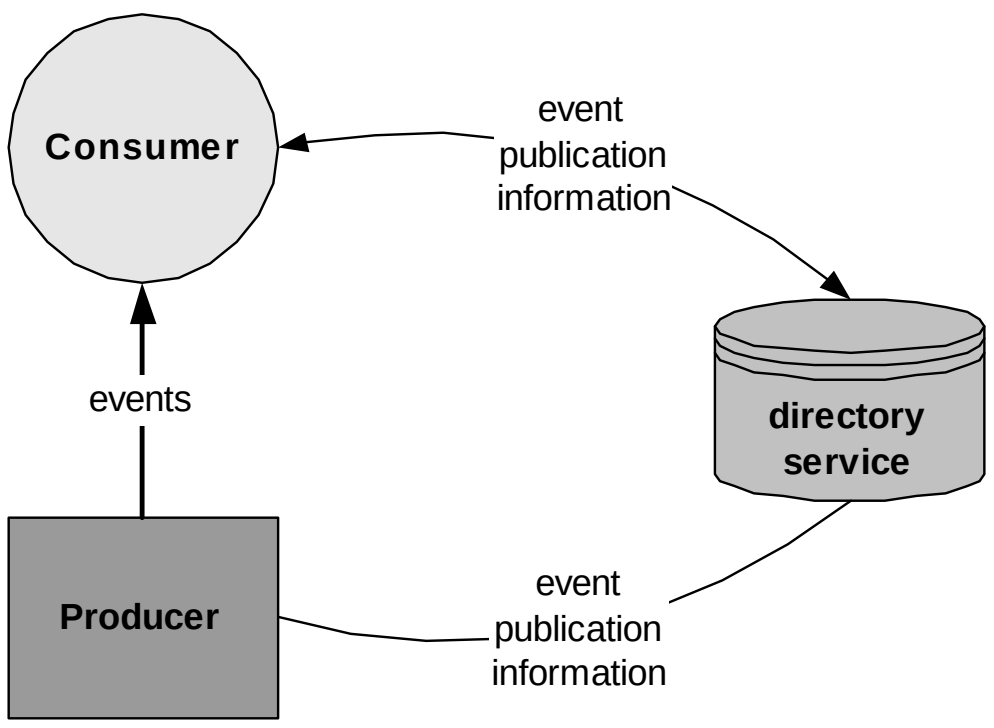

**Figure 2: GMA Components**

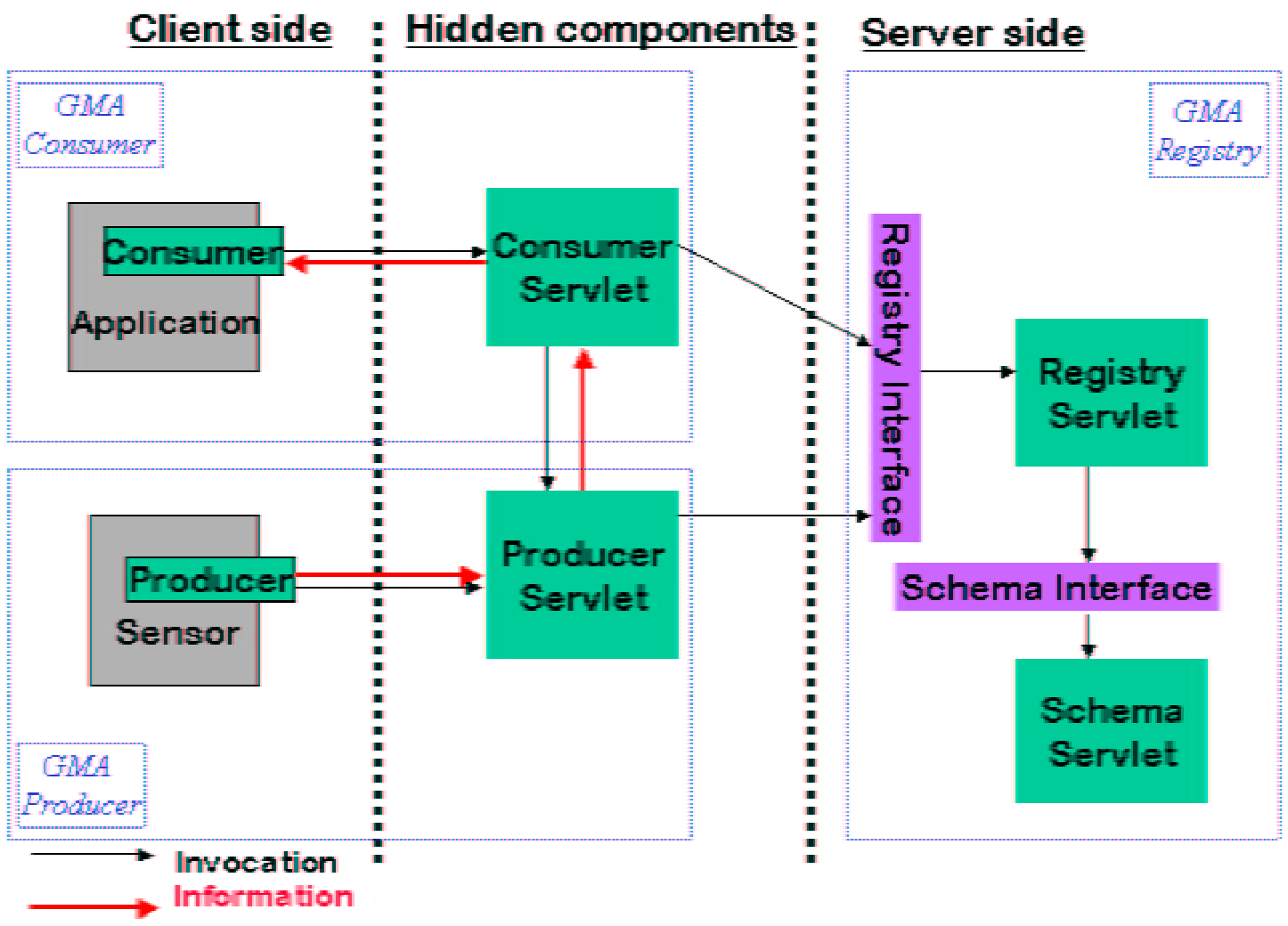

**Figure 3: R-GMA Components**

Figure 3 illustrates the R-GMA components and their mapping to GMA. In R-GMA, to register with a Registry, a Producer advertises a table name and the row(s) of a table to the Registry. The Producer module communicates with a ProducerServlet, which registers the information to the RDBMS in the Registry. The RDBMS holds the information for all the Producers (the registered table name, the identity, and the values of those fixed attributes) and the descriptions of each Producer's tables. Consumers can issue SQL queries against a set of supported tables. The ConsumerServlet consults the Registry to find suitable Producers. Then the ConsumerServlet acting on behalf of the Consumer issues new queries to the located Producers to request and return the data to the Consumer. The ProducerServlet and ConsumerServlet are usually distributed and may run on machines remote from where the Producer or Consumer is located.

## 2.3 Hawkeye

Hawkeye [10] is a tool developed by the Condor group and designed to automate problem detection (e.g., high CPU load, high network traffic, or resource failure) and software maintenance within a distributed system. Its underlying infrastructure builds on the Condor [13] and ClassAd [18] technologies. The main use case that Hawkeye was built to address is that of being able to offer warnings (e.g., high CPU load, low disk space, or resource failure) using *Trigger ClassAds*. It also allows for easier software maintenance within a Pool.

Hawkeye involves two fundamental ideas: its use of the Condor *ClassAd* Language to identify resources in a Pool; and *ClassAd Matchmaking* [18][19] to execute jobs based on attribute values of resources to identify problems in a pool. A *ClassAd* is a set of attribute/value pairs (e.g., "operating system" and "Linux"). The Manager performs *ClassAd Matchmaking* between a *Trigger ClassAd*, submitted by a client, and all *Startd ClassAds*. A *Trigger ClassAd* specifies an event and a job to execute if the event occurs. For example, consider the case in which a Trigger ClassAd specifies an event in which the CPU load is greater than 50 and a job that will kill Netscape on the matched machine; if any machine advertises a Startd ClassAd with a CPU load value of greater than 50, the Manager will kill that machine's Netscape process.

The architecture of Hawkeye comprises four major components: Pool, Manager, Monitoring Agent, and Module (Figure 4). The components are organized in a four-level hierarchical structure. A Pool is a set of computers, in which one computer serves as the Manager and the remaining computers serve as Monitoring Agents. A Manager is the head computer in the Pool that collects and stores (in an indexed resident database) monitoring information from each Agent registered to it. It is also the central target for queries about the status of any Pool member. A Monitoring Agent is a distributed information service component that collects ClassAds from each of its Modules and then integrates them into a single *Startd ClassAd*. At fixed intervals, the Agent sends the *Startd ClassAd* to its registered Manager. An Agent can also directly answer queries about a particular Module; however, the client must first consult the Manager for the Agent's IP-address. A Module is simply a sensor that advertises resource information in a *ClassAd* format.

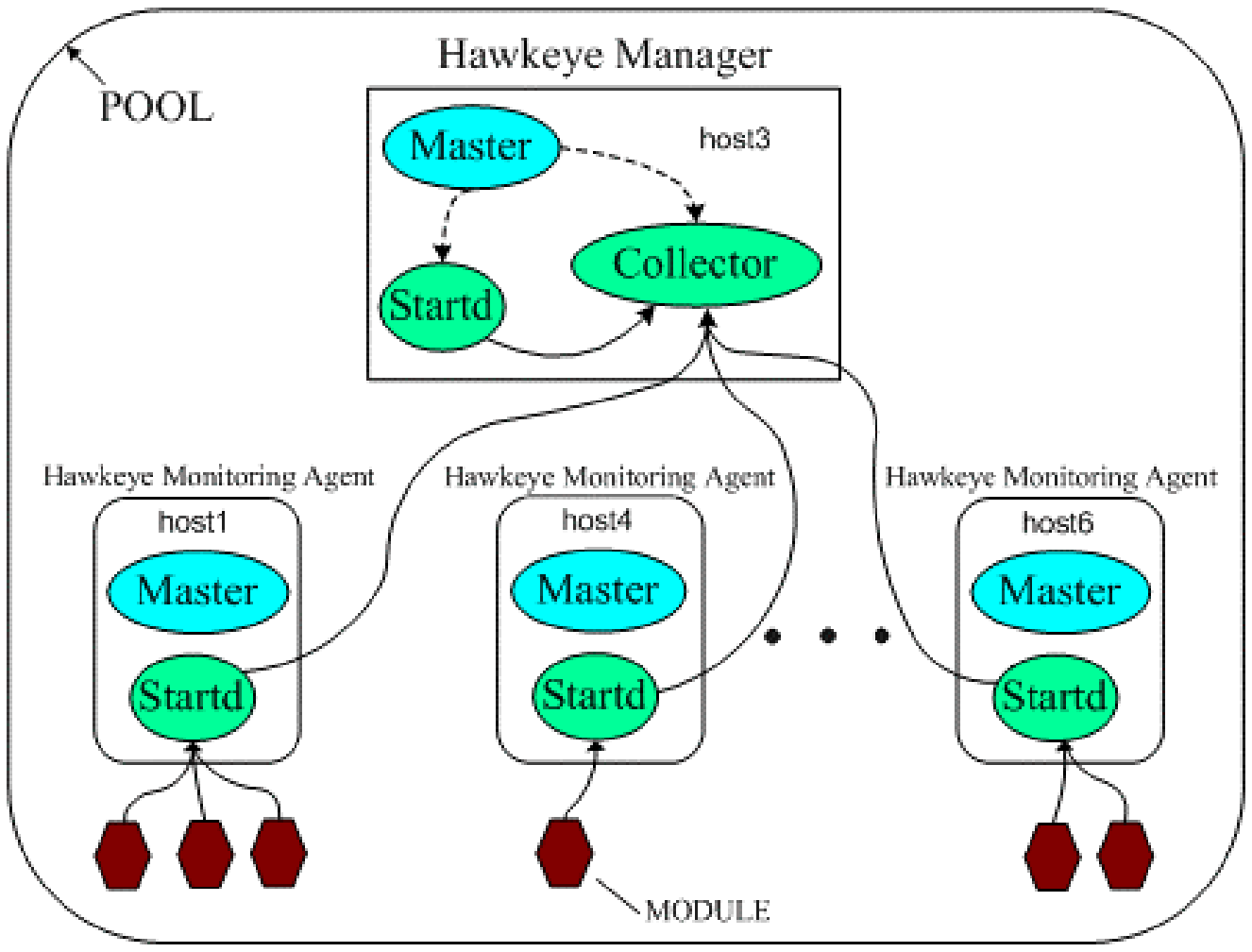

**Figure4: Summary of the Hawkeye architecture along with the basic function of each node.**

## 2.4 Components Mapping

In addition to examining the individual performance of MDS, R-GMA and Hawkeye under various environments, we compare the behavior between the systems. To facilitate this comparison, we map the functional components of the services to one another.

**Table 1: Component Mapping**

| | MDS | R-GMA | Hawkeye |
|---|---|---|---|
| **Information Collector** | Information Provider | Producer | Module |
| **Information Server** | GRIS | ProducerServlet | Agent |
| **Aggregate Information Server** | GIIS | None | Manager |
| **Directory Server** | GIIS | Registry | Manager |

Table 1 shows the mapping selections defined for this comparison. At the lowest level, we have the *Information Collector*, which is equivalent to a sensor or other program that generates a piece of data. This maps to the MDS information provider, an R-GMA Producer, or a Hawkeye Module. At the resource level, these systems gather together the data from several information collectors into a component we term the *Information Server*. Examples of this are the MDS GRIS, an R-GMA ProducerServlet, and a Hawkeye Agent. Some systems allow data to be aggregated from a set of resources; we term this an *Aggregate Information Server*. Examples are the MDS GIIS and the Hawkeye Manager. This component could easily be built for R-GMA by using a composite Consumer/Producer that registered with the data streams of a number of Producers, and server the data in an aggregated form. The fourth component of the systems is a *Directory Server*. This role is played by the GIIS in MDS, the Registry in R-GMA and the Manager in Hawkeye.

## 3. Experimental Results and Evaluation

In this section we briefly describe the experimental setup and the performance metrics. We then analyze and compare the results of the experiments.

The goal of these experiments was to test the scalability of the components of the information systems. Since each system had a different architecture and different design principles, not all of the comparisons are direct. We considered the following:

- How does the performance of an information server scale with the number of users?
- How does the performance of a directory server scale with the number of users?
- How does the performance of an information server scale with the amount of data it contains (the number of information collectors)?
- How does an aggregate information server scale with the number of information servers it is aggregating?

By evaluating these questions we can make suggestions for deployment or future updates to the systems themselves.

### 3.1 Experimental Setup

The experiments were run on the Lucky testbed at Argonne National Laboratory (ANL). The Lucky testbed includes seven Linux machines with hostnames lucky{0,1,3,..,7}.mcs.anl and a shared file system on a 100 Mbps LAN. Each machine is equipped with two 1133 MHz Intel PIII CPUs (with a 512 KB cache per CPU) and 512 MB main memory and runs a Linux kernel 2.4.10. During our experiments, MDS 2.1, R-GMA 1.18 (9/2002) and Hawkeye 0.1.4 were deployed in the Lucky testbed.

The values reported are the average over all the values recorded during a 10-minute time span. We used Ganglia [8] to collect the performance data at five-second intervals. All requests to the servers in MDS, R-GMA, and Hawkeye occurred with a one-second *wait period*. That is, after a user queried a service component and received a response, the user waited one second before sending its next query. Note this does not mean that queries were sent once a second, rather, this is equivalent to blocking sends with a 1-second wait in between each.

We simulated users by running individual user processes (scripts) at a cluster of 20 client machines located at the University of Chicago (UC). Fifteen of them were equipped with a 1208 MHz uniprocessor and 248 MB RAM, while the rest had a slightly slower CPU (but at least 756 MHz), also with 248 MB RAM. When simulating users, we evenly divide the number of simulated users by the number of machines to balance the load, with a maximum of 50 users per machine. In the case of R-GMA, we could simulate only 120 Consumers per ConsumerServlet because of a limit of 128 in the table size and the shared filespace on this system (so a single table shared among all UC hosts). This is not an issue with R-GMA, but of our testing environment. Because of this situation, for the R-GMA experiments we also simulated a larger number of users on the Lucky testbed with a ConsumerServlet running on each Lucky node.

### 3.2 Performance Metrics

*Throughput* and *response time* were the primary metrics in our study. We define *throughput* as the average number of requests (or queries) processed by a service component per second. The *response time* denotes the average amount of time (in seconds) required for a service component to handle a request sent from a user.

We also used two load metrics for the experiments, CPU-load (*load)* and a one-minute load average (*load1).* *Load* indicates the percentage of the CPU cycles spent in user mode and system mode, we measured by averaging the sum of cpu_user and cpu_system recorded by Ganglia. The metric *load1* is the average number of processes in the ready queue waiting to run over the last minute measured by the Ganglia metric "load_one". *Load* may be high while *load1* is low when a machine is running a small number of compute intensive applications. *Load* may be low, while *load1* is high, if the same machine is trying to run a large number of applications that are blocking on I/O.

## 3.3 Experiment Set 1 –Information Server Scalability (with users)

The information server is the most heavily accessed component in a monitoring and information service. We evaluated the performance of the information servers for three systems—the MDS GRIS, the R-GMA ProducerServlet, and the Hawkeye Agent—when a large number of users accessed it concurrently.

To study the MDS GRIS, we ran a GRIS and 10 information providers at *lucky7* with two different configurations: the information provider data always in cache and the data never in cache. We simulated up to 600 users making concurrent queries.

For the study of the ProducerServlet in R-GMA, we ran a ProducerServlet at *lucky3* with 10 local Producers, and a Registry on *lucky1*. We simulated users (Consumers) in the two ways as described in Section 3.1: with up to 100 simulated users on the UC nodes reporting to a single ConsumerServlet, and with up to 600 users on the Lucky nodes, with each node running its own ConsumerServlet.

To study the performance of the Hawkeye Agent, we ran an Agent on *lucky4* and the Hawkeye Manager on *lucky3*. We simulated up to 600 users to concurrently query the Agent from the UC nodes.

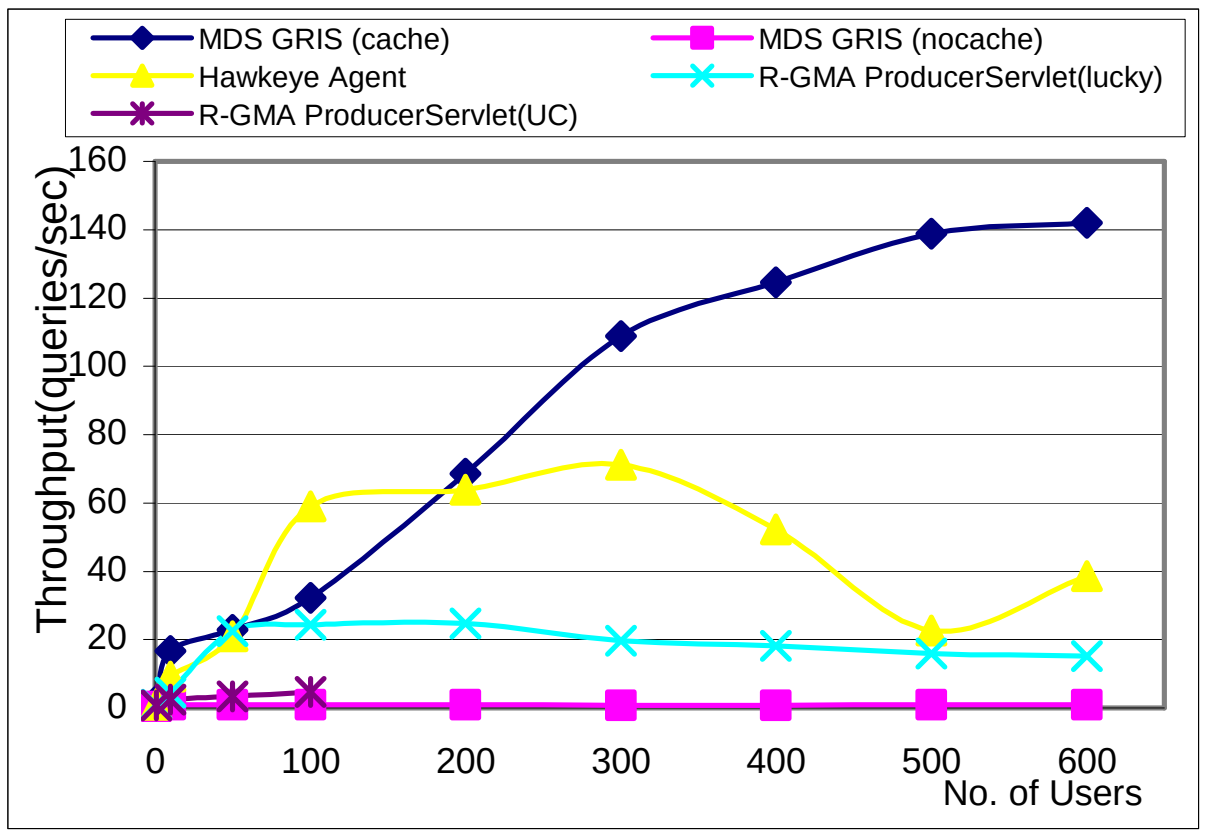

**Figure 5: Information Server Throughput vs. No. of Concurrent Users**

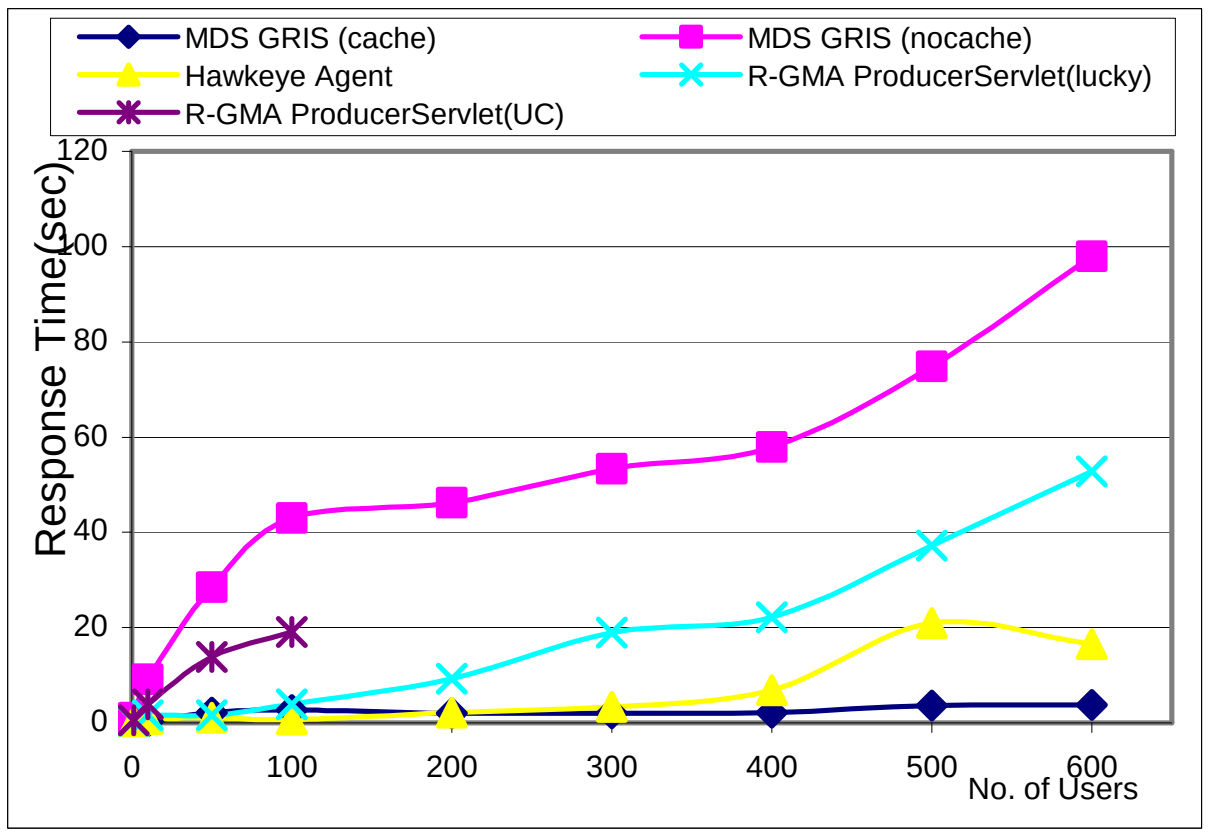

**Figure 6: Information Server Response Time vs. No. of Concurrent Users**

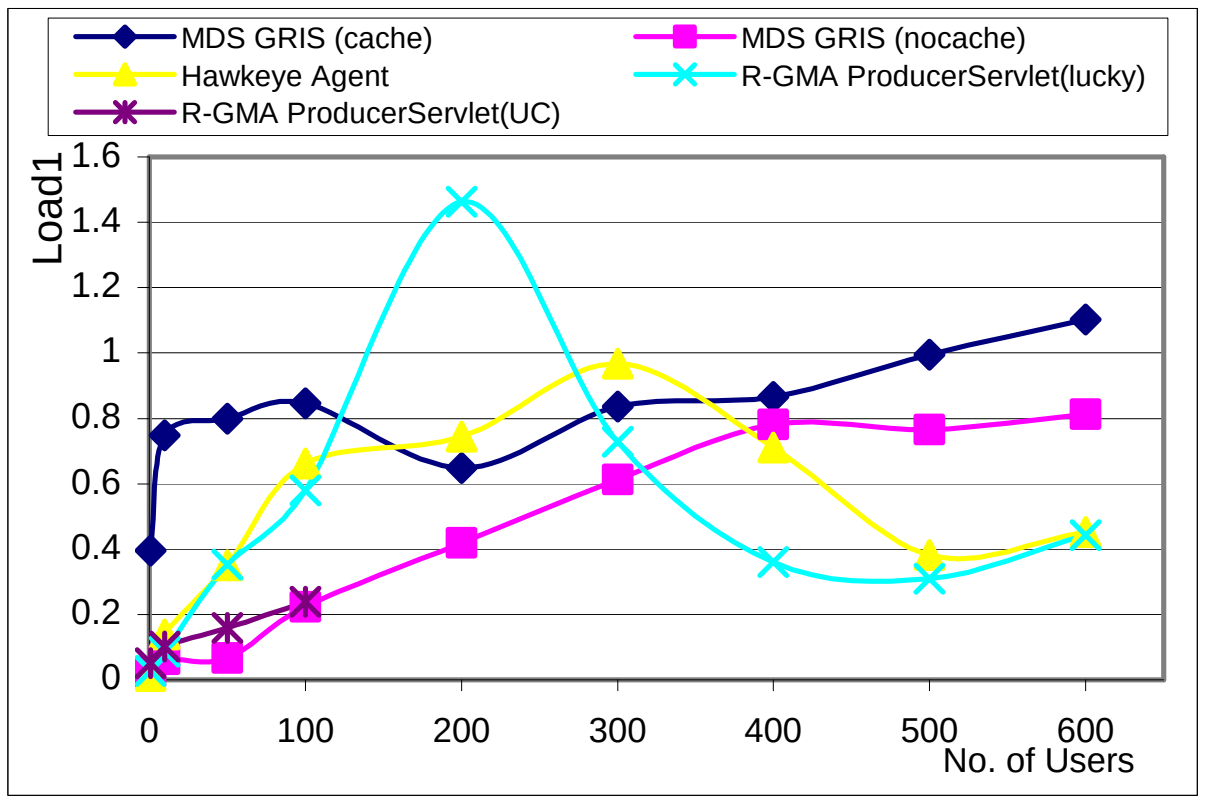

**Figure 7: Information Server Load1 vs. No. of Concurrent Users**

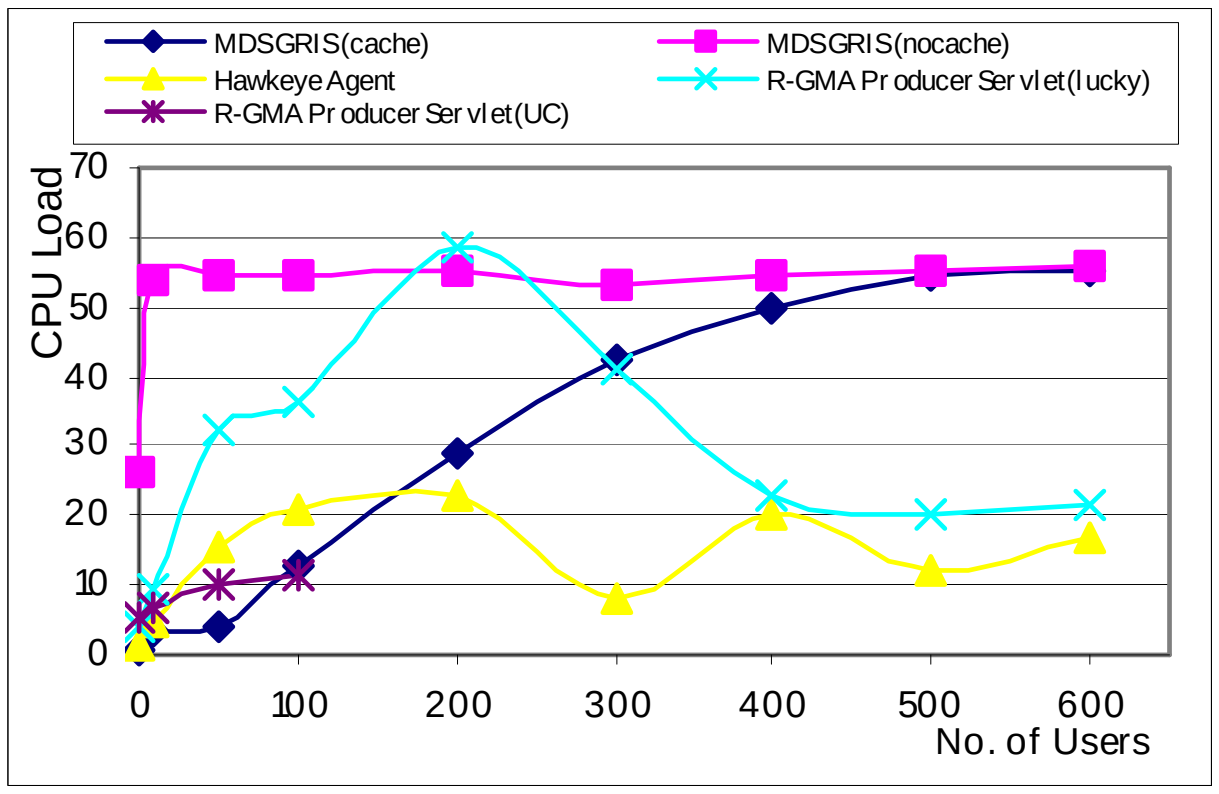

**Figure 8: Information Server CPU Load vs. No. of Concurrent Users**

Figures 5–8 show the performance of the three information servers including two different scenarios for R-GMA and two for MDS. The MDS GRIS throughput has a near linear relationship with the number of concurrent users if the data is in cache. Its throughput, however, does not exceed 2 queries per second when the data is not in cache, suggesting that data caching is very important to support a large number of users. We observe throughput thresholds for the information servers that did not cache data. We believe this is because the network on the server side can no longer handle the traffic from the queries, which limits the number of concurrent queries presented to the information server, and so the throughput does not rise above a set level.

With respect to response time behavior, the GRIS results show stable performance (approximately 4 seconds per query) for 50 concurrent users or more if the GRIS has data in cache. Otherwise the performance is much worse. For the R-GMA ProducerServlet, the response time grows almost linearly with the number of users, indicating that for larger number of users the system should be configured to have multiple ProducerServlets for the same information to allow better scaling. The Hawkeye Agent had under a 10 second response time per query until 500 users are tested, quite likely because the Hawkeye Agent does not hold the indexed resident database as the Hawkeye Manager and it has to retrieve new information for each query.

From Figures 7–8, we also observe that the machines hosting the three information servers experienced a higher load with the increasing number of users below the load threshold. The load decreased sharply after the threshold for both the Hawkeye Agent and the R-GMA ProducerServlet indicating that a large percentage of the processes were blocked waiting for resources.

These experiments show that caching can significantly improve performance of the information server; caching is particularly desirable if one wishes the server to scale well with an increasing number of users. In addition, when setting up an information server, care should be taken to make sure the server is on a well-connected machine, since network behavior plays a larger role than expected. If this is not an option, thought should be given to duplicating the server if more than 200 users are expected to query it.

### 3.4 Experiment Set 2 –Directory Server Scalability

The second functionality we tested was the scalability of the directory server with the number of users. In particular, we examined the performance of the MDS GIIS, the R-GMA Registry and the Hawkeye Manager.

We configured the testbed for the MDS GIIS to run a GIIS on *lucky0* with a GRIS and 10 information providers on each of *lucky3-7* registered to it. We simulated up to 600 users on the UC client nodes. To analyze only the directory functionality of the GIIS and not its information serving capacity, we set the *cachettl* (cache element time to live) parameter to a very large value so that the data was always in the cache.

For the Hawkeye Manager experiment, we ran the Manager at *lucky3* and up to 600 users concurrently querying the Manager at the UC client nodes. There were 6 Agents (one on each Lucky node) each running 11 default Modules.

For R-GMA we ran the Registry at *lucky1* and one ProducerServlet on each of five other Lucky nodes with 10 local Producers. Again, we simulated users in two ways, up to 100 with a local ConsumerServlet on the UC nodes and up to 600 users on the Lucky nodes.

Figures 9–12 illustrate that both the MDS GIIS and Hawkeye Manager present good scalability with respect to the number of users, while R-GMA had slightly less. After approximately 10 users, the load on the Hawkeye Manager rapidly increases, reaching a maximum *load1* average of .45 at about 200 users. At 400 users, the load begins to decrease gradually. We believe this decrease occurs because the network on the server side can no longer handle the traffic from the queries.

For MDS, the throughput goes up quickly when the number of users is smaller than 100, then saturates at approximately 200 users. We believe the primary cause of this threshold is similar as that for Hawkeye Manager. The response time remains relatively small (less than 2 seconds) even as the number of users increases (up to 600). However, we observe the load of GIIS (Figure 12) is nearly twice as bad as Hawkeye Manager when the number of users is large. This is likely due to better efficiency in the indexed resident database used in the Manager than seen in the LDAP backend used in GIIS.

The R-GMA Registry presents a lower throughput and higher load. The reduction in throughput may be due to the fact that R-GMA is based on Java, so must spawn additional threads to handle the user queries. The results also show that there is little difference between the performances of R-GMA's Registry when accessed by two different kinds of simulated Consumers (remotely from UC and locally from Lucky nodes), which suggests that the effect of networking constraint is less significant than the resource contention at the Registry.

From these experiments we see again that, because of the network contention issues, the placement of a directory server on a highly connected machine will play a large role in the scalability as the number of users grows. In addition, because significant loads are seen even with only a few users, it will be important that this service be run on a dedicated machine, or that it be duplicated as the number of users grows.

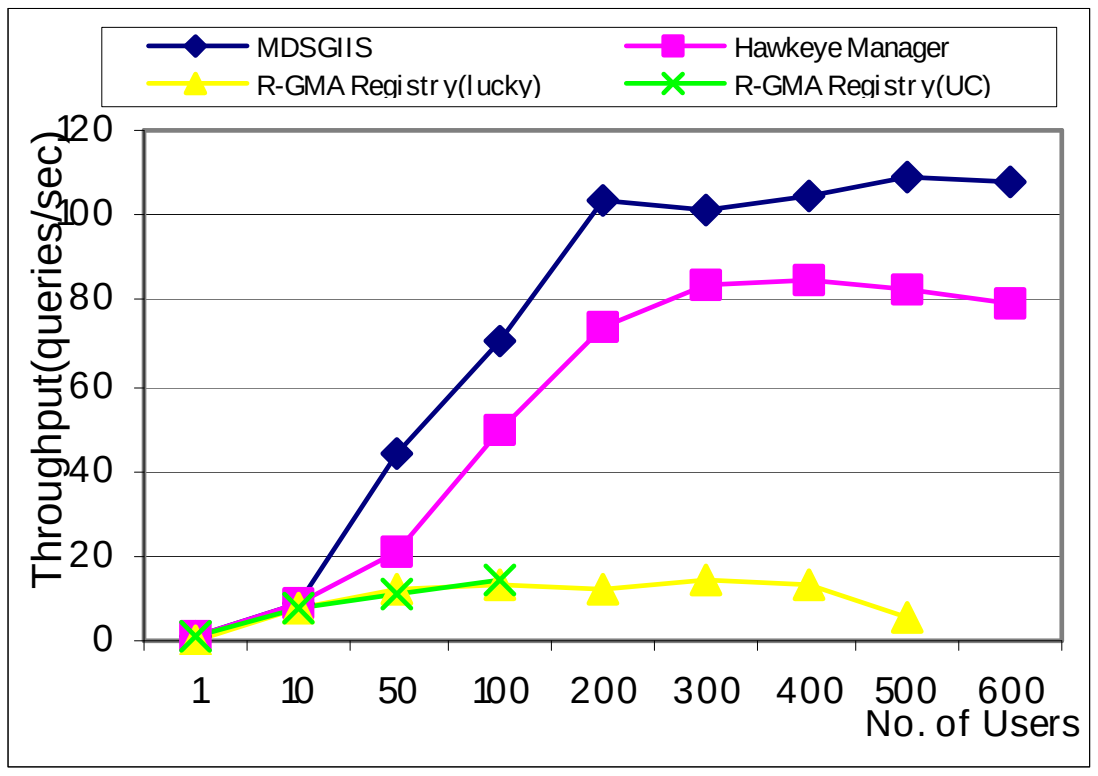

**Figure 9: Directory Servers Throughput vs. No. of Concurrent Users**

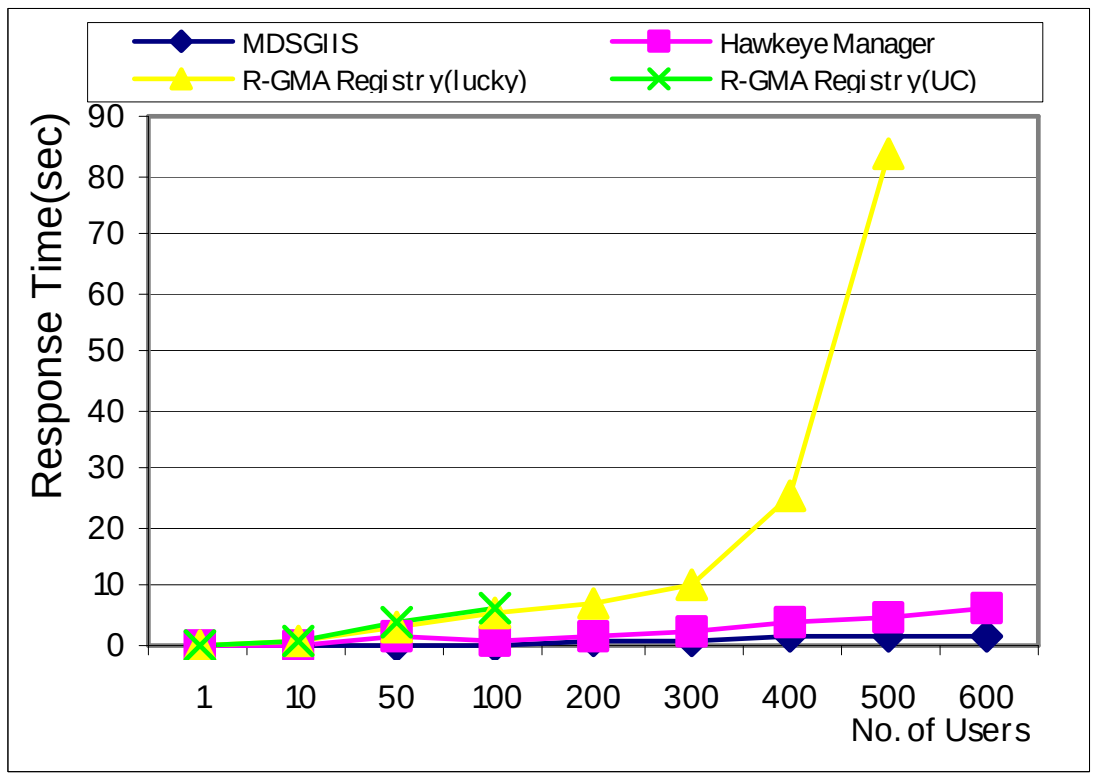

**Figure 10: Directory Servers Response Time vs. No. of Concurrent Users**

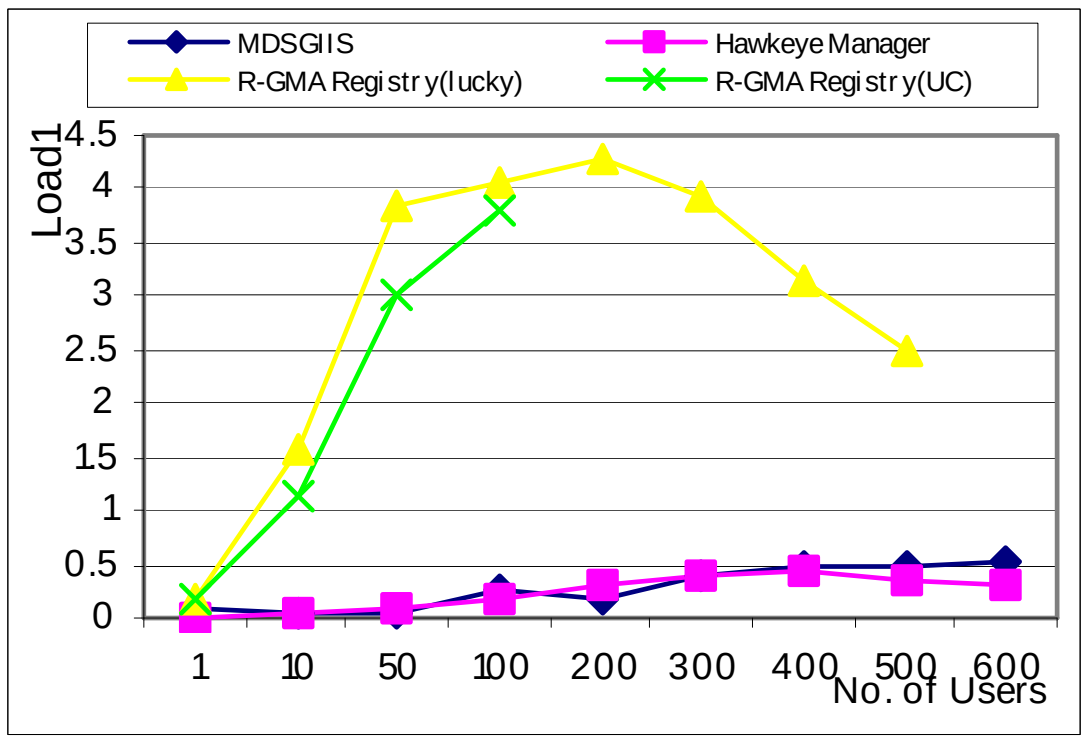

**Figure 11: Directory Servers Host Load1 vs. No. of Concurrent Users**

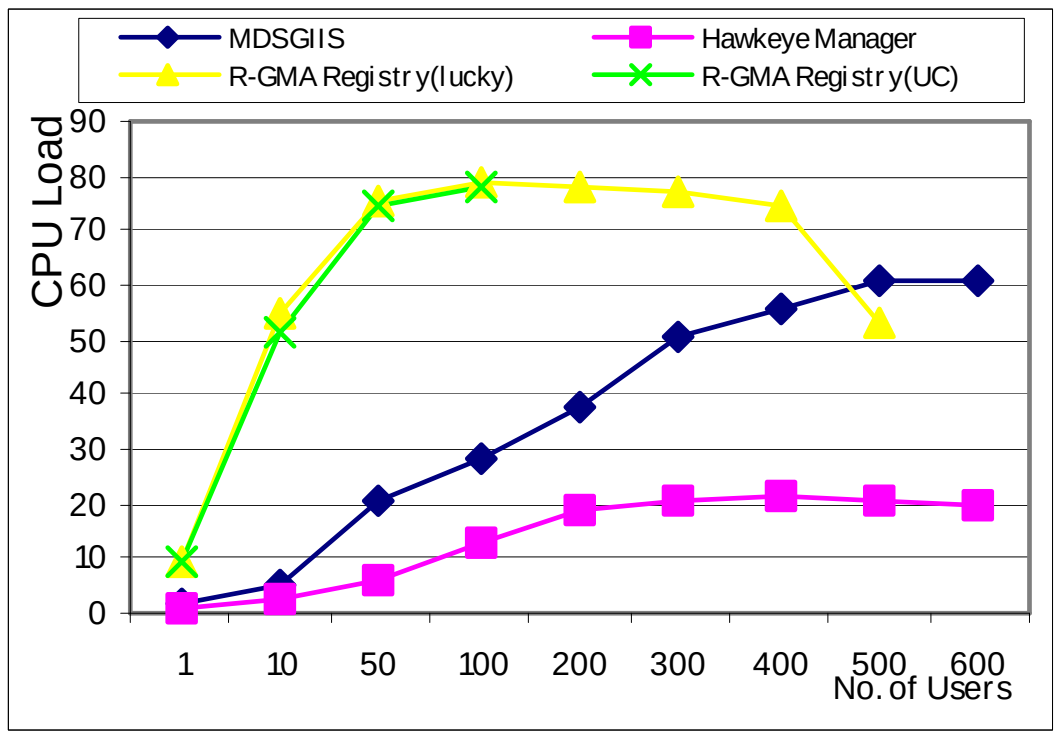

**Figure 12: Directory Servers Host CPU Load vs. No. of Concurrent Users**

## 3.5 Experiment Set 3 –Information Server Scalability (with information collectors)

Our next set of experiments evaluated how the performance of an information server scaled with the amount of data it contained, specifically the MDS GRIS and its information providers, the Hawkeye Agent and its Modules, and the R-GMA ProducerServlet and Producers. Currently, a default installation of the MDS has 10 information providers, and Hawkeye uses 11 Modules in a standard install, while R-GMA currently has a flexible configuration for the initial set of Producers. The importance of this study lies in the fact that new information collectors will be added to the monitoring and information services in the near future, and so information servers must be able to scale. For example, other monitoring systems (such as WatchTower [12]) can publish as many as 2,000 individual pieces of information from a single machine.

For the MDS experiments, we modified the default memory information provider and added copies of the new version to simulate the expanded information providers. The GRIS was located at *lucky7* and 10 concurrent users sent queries to it for the information from all the information providers for up to 90 information providers reporting to the same GRIS. We also simulated two different cases of caching for the GRIS: the data always in the GRIS cache and never in the GRIS cache.

For Hawkeye, we varied the total number of Modules running on each pool member from the 11 default Modules to 90 using multiple instances of the "vmstat" Module. Once the Modules were running on each pool member, 10 concurrent users queried the Manager. The maximum number of Modules currently able to register to an Agent was 98, adding another Module caused the Startd to crash.

For R-GMA, we ran the ProducerServlet at *lucky3* and the Registry on *lucky1* with up to 90 Producers. We used 10 Consumers from the UC client nodes with a ConsumerServlet at UC. We queried the ProducerServlet directly.

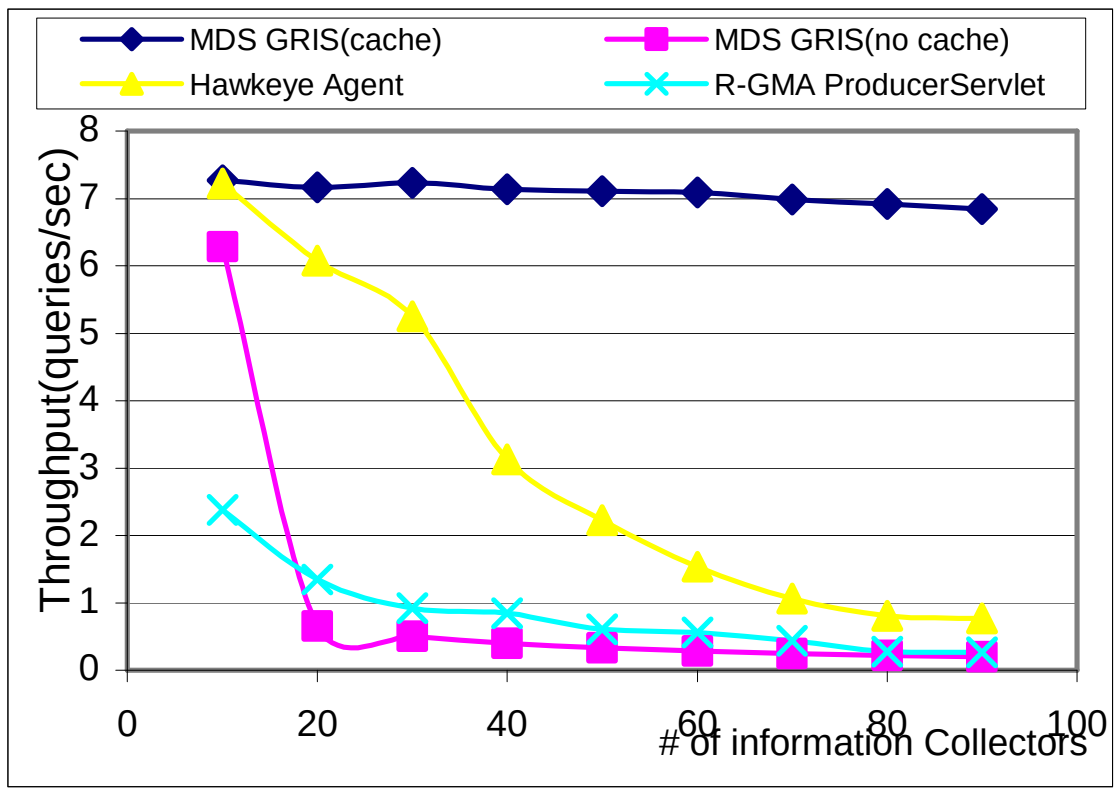

**Figure 13: Information Server Throughput vs. No. of Information Collectors**

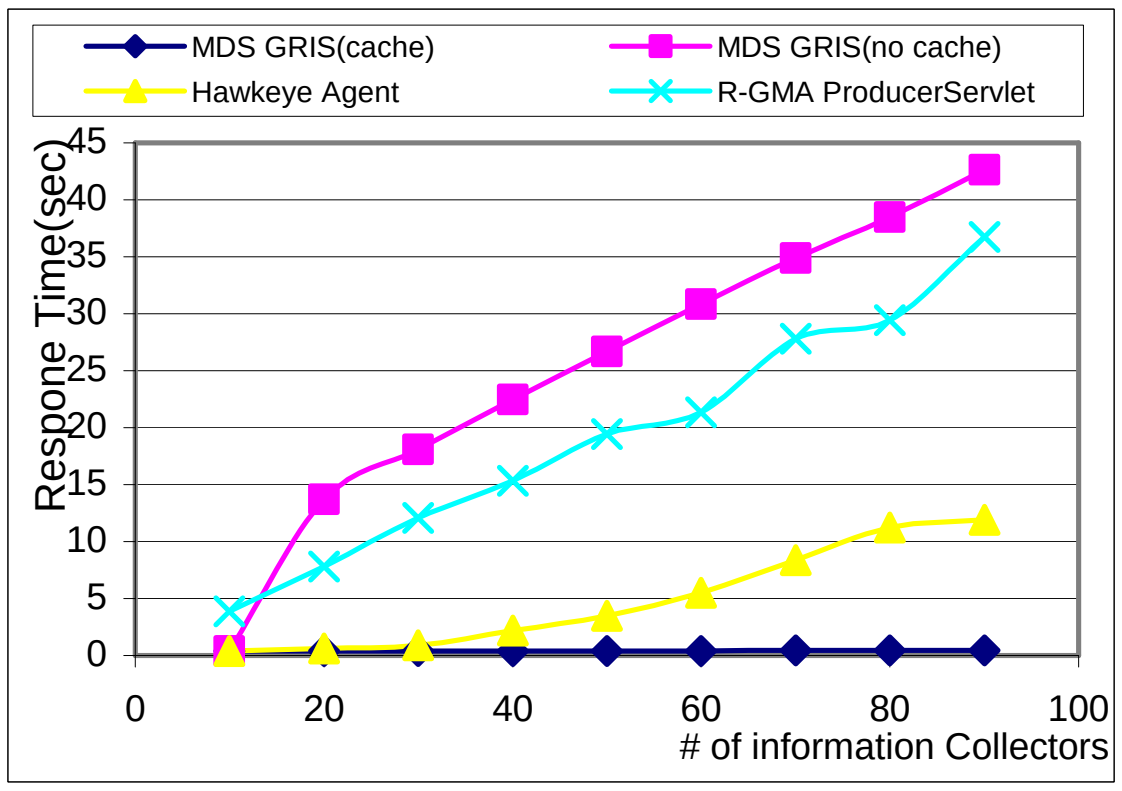

**Figure 14: Information Server Response Time vs. No. of Information Collectors**

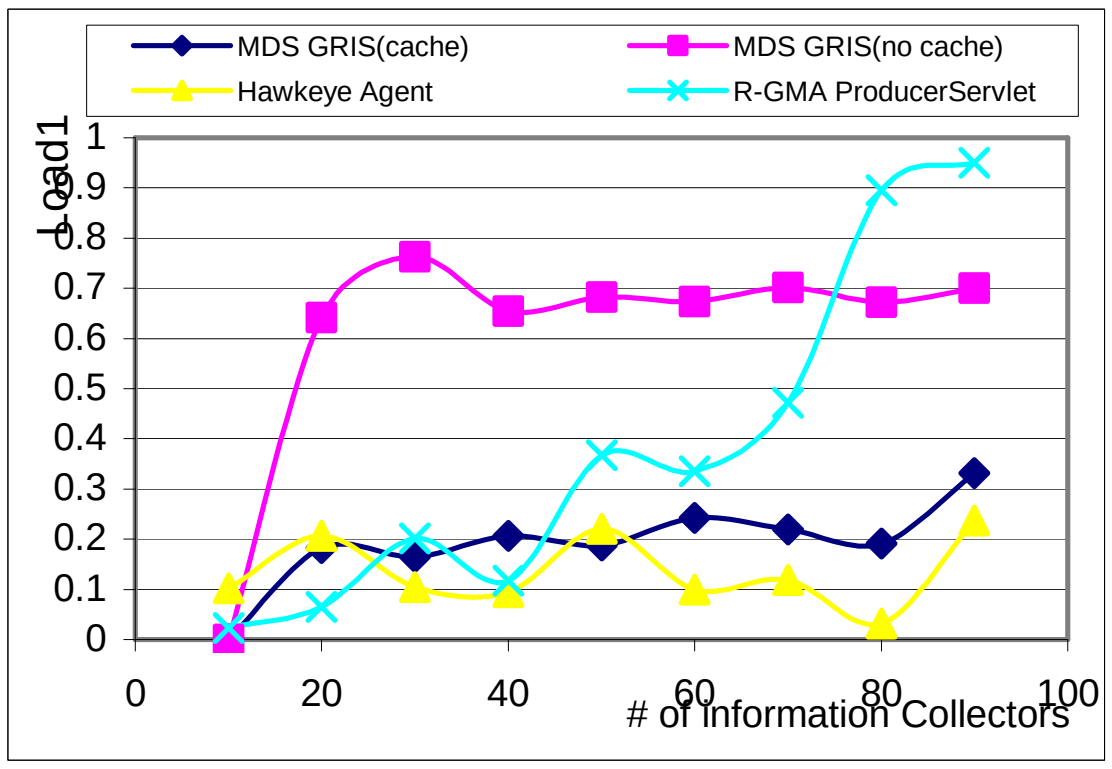

**Figure 15: Information Server Host Load1 vs. No. of Information Collectors**

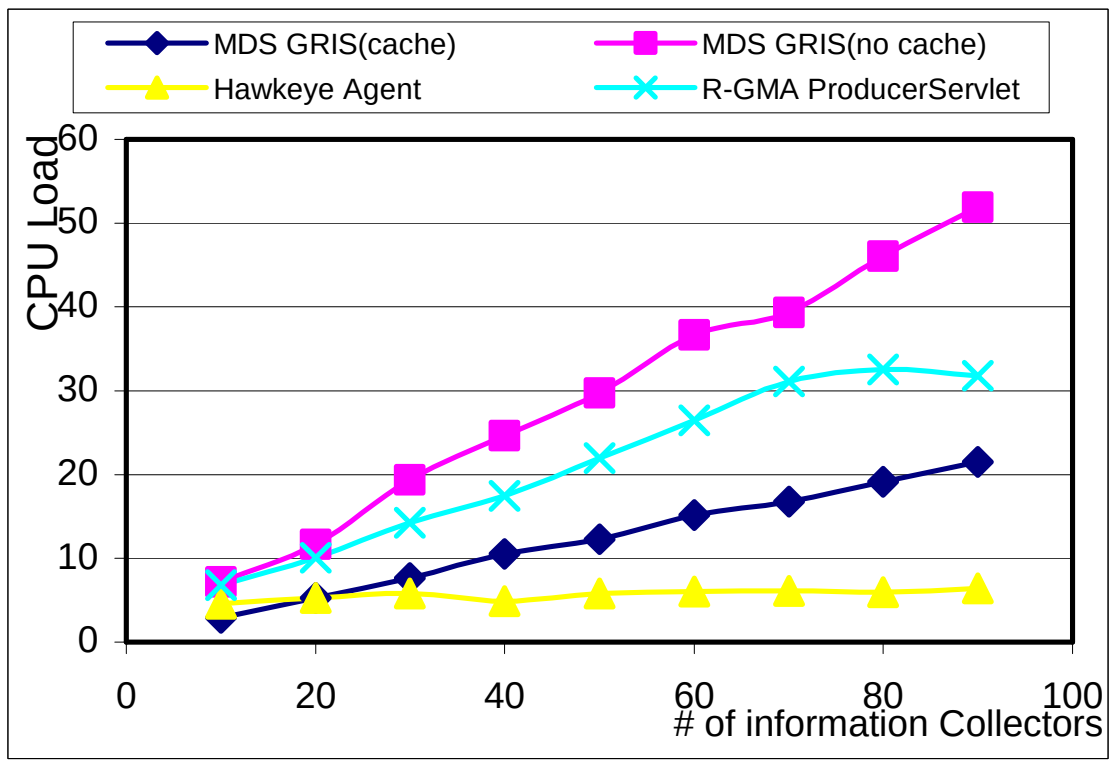

**Figure 16: Information Server Host CPU Load vs. No. of Information Collectors**

In Figures 13–16, we observe that the performance of all three information servers degrades when the number of information providers grows. If the data is always in the GRIS cache, however, it can still achieve a throughput of 7 queries per second with a less than 1-second response time for 90 information collectors. The Hawkeye Agent and the R-GMA ProducerServlet have throughput levels of less than 1 query per second and over 10-second response times. One possible reason for this behavior is that the caching mechanism in the GRIS allows it to not re-fetch the data from the information collectors.

Our analysis shows that too many information collectors can become the performance bottleneck but that caching data at the information server can be an effective solution. Alternatively, a large number of information collectors can register to more instances of information servers with each handling a subset of the collectors. How to coordinate these instances needs further investigation.

## 3.6 Experiment Set 4 –Aggregate Information Server Scalability

Our fourth set of experiments examines the aggregation of data, specifically the scalability of the MDS GIIS varying the number of GRIS, and the Hawkeye Manager varying the number of Agents. R-GMA currently has no aggregate information server, but one could easily be built using a composite Consumer/Producer that registered with the data streams of a number of Producers, and served the data in an aggregated form. Since this component is not yet part of the standard distribution, we compare only the MDS and Hawkeye for this experiment.

For the MDS, we simulated extra GRIS by running multiple instances at each Lucky node, except on *lucky0* where the GIIS ran. We were able to simulate up to 500 GRIS; adding more caused the GIIS to crash. During the experiment, 10 users from a UC node concurrently sent queries to the GIIS for 10 minutes. We tested two kinds of queries: the first queried for all of the data available from each of the registered GRIS, and the second asked for only a portion of the data from each registered GRIS.

For Hawkeye, we simulated the large number of Agents (computers) in a pool by using the "hawkeye_advertise" command to send Startd ClassAds at 30-second intervals to the collector machine. Once the simulated machines were running, ten concurrent users queried the Manager for information. The Manager searched through all of the ClassAds (one for each of the simulated machines) to try to match on a constraint that was not met by any machine, simulating the worst-case scenario.

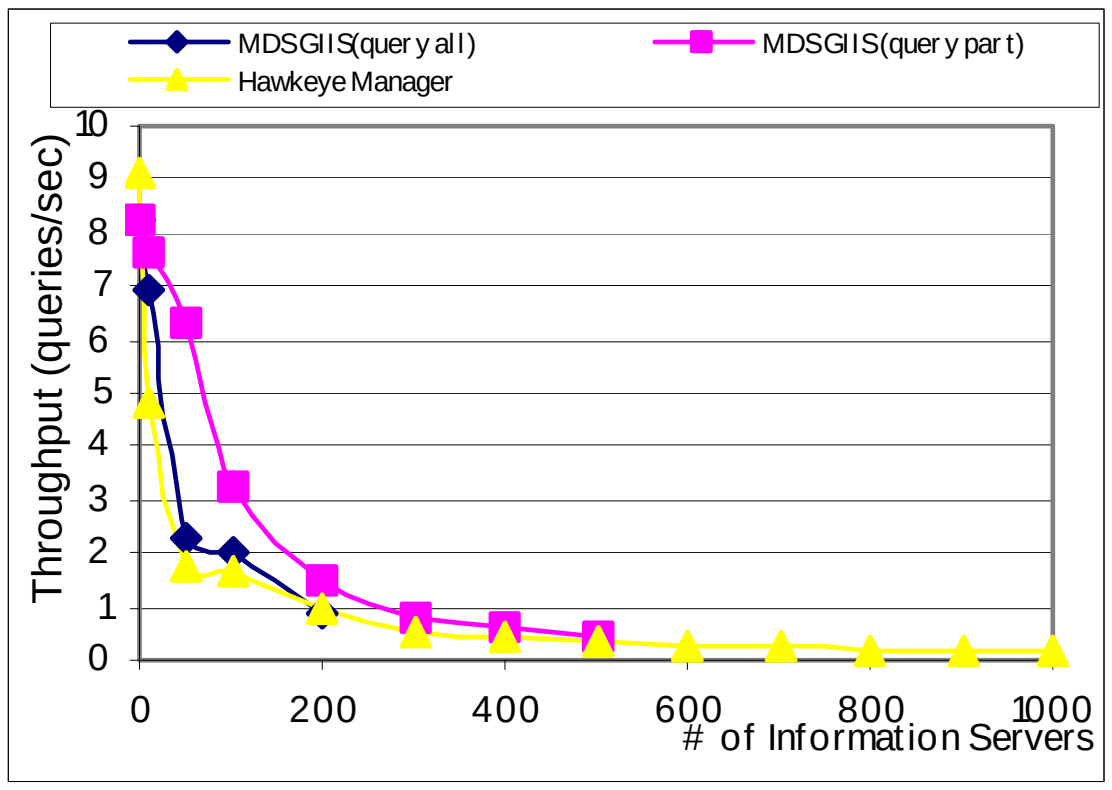

**Figure 17: Aggregate Information Server Throughput vs. No. of Information Servers**

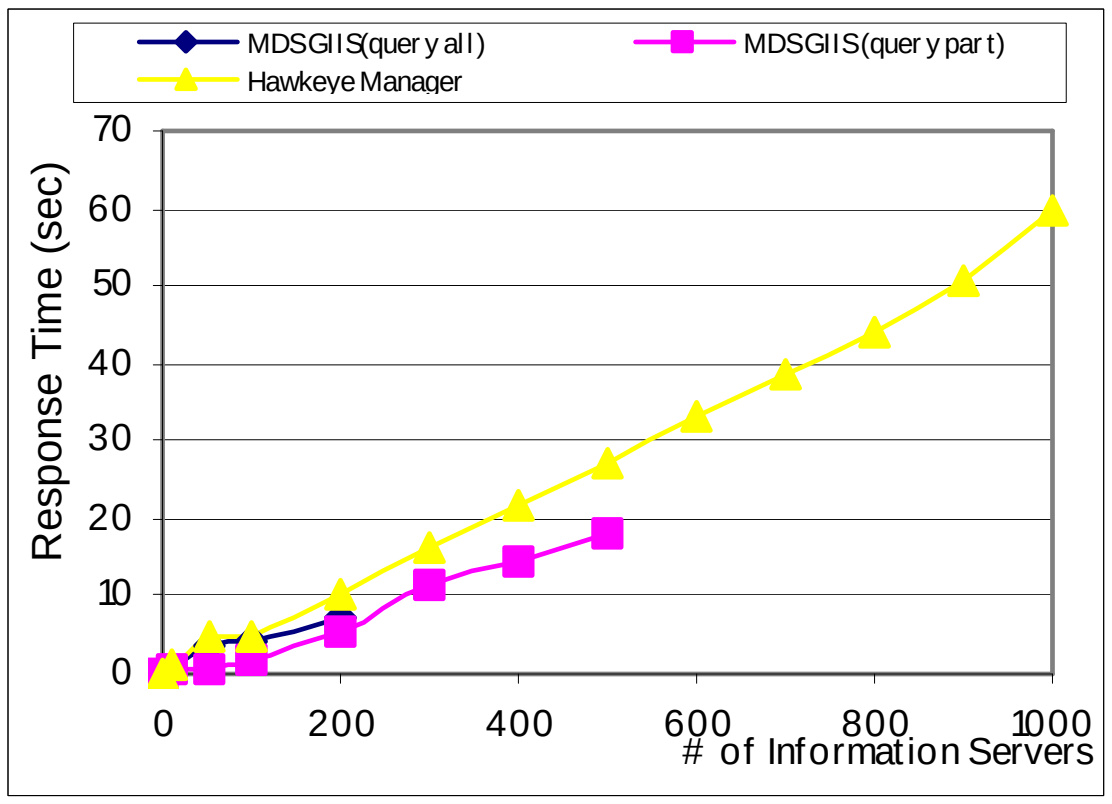

**Figure 18: Aggregate Information Server Response Time vs. No. of Information Servers**

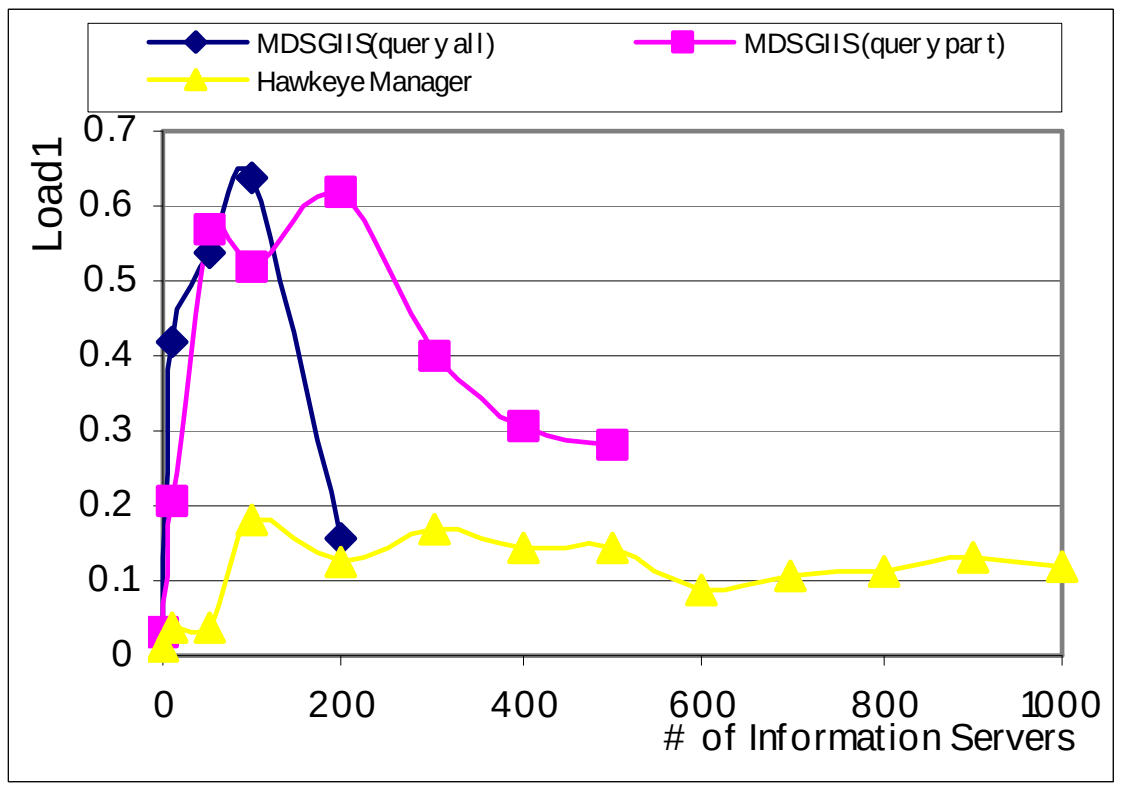

**Figure 19: Aggregate Information Server Host Load1 vs. No. of Information Servers**

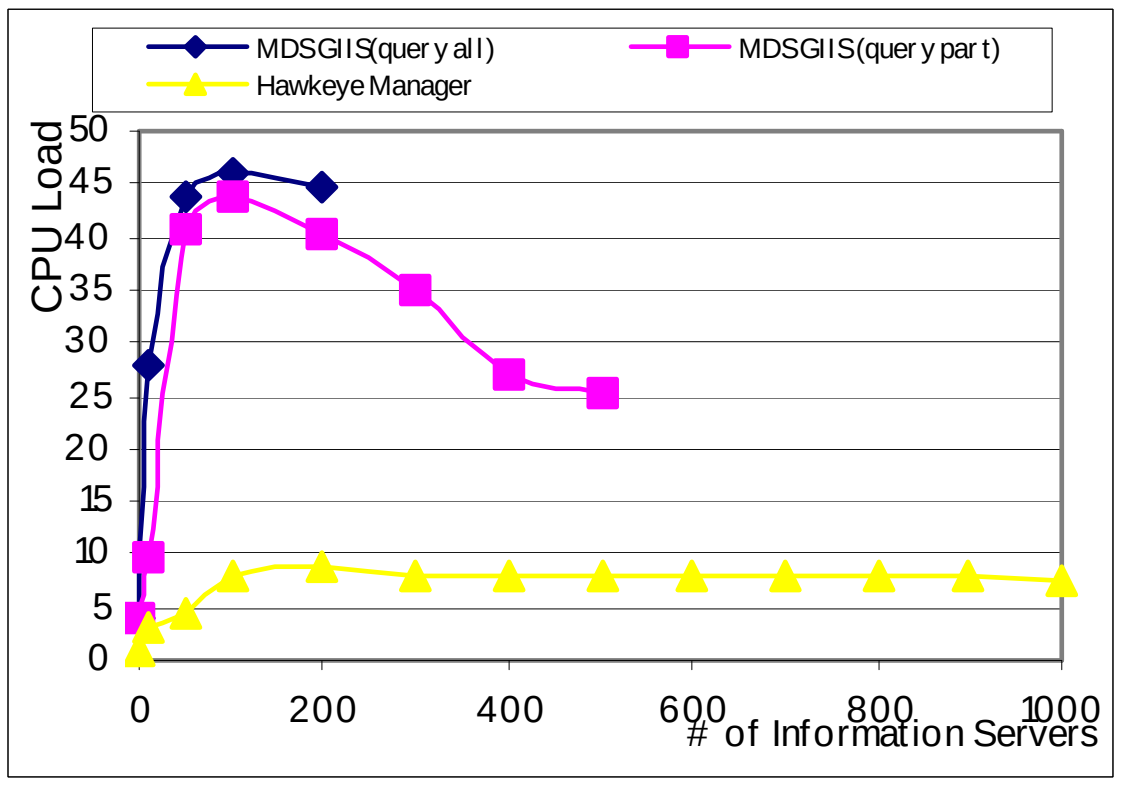

**Figure 20: Aggregate Information Server Host CPU Load vs. No. of Information Servers**

From the experiments results shown in Figure17-20, we observe a large degredation in both throughput and response time by the MDS GIIS and Hawkeye Manager.

For the MDS GIIS, because of the software constraints, we were able to simulate only up to 200 registered GRIS if all the users queried GIIS for the information from all the registered GRIS and up to 500 GRIS if all the users queried GIIS for the information from part of the registered GRIS. In each case, the machine hosting the GIIS reaches a maximum load at around 100 registered GRIS.

For the Hawkeye Manager, after approximately 10 simulated computers, the load on the Manager rapidly increases until about 200 computers. After 200, the load begins to decrease slightly but remains relatively constant. The reason that the load does not continue to increase is that the number of incoming ClassAds to the Manager is high, which once again overloads the network at the server-end.

The experimental results suggest that no current aggregate information server is capable of aggregating information servers when there are more than 100 of them. To achieve a higher scalability for an aggregate information server, a multi-layer architecture in which each middle-level aggregate information server manages a subset of information servers should be examined.

### 3.7 Review of MDS, R-GMA and Hawkeye

MDS and R-GMA share many similarities; in particular, both provide solutions for a Grid information system and are driven by basic properties of the Grid environment. Hawkeye, too, has the same functional components (i.e., an information collector, an information server, and a directory), although it focuses on monitoring a single pool. All three services describe the monitoring information uniformly using a global schema (e.g., LDAP in MDS, Relational Data Model in R-GMA, and ClassAds in Hawkeye).

The three services, however, have several key differences in design goals, architecture, and underlying technologies. For example, the MDS GIIS provides a simple registration mechanism (where a GIIS registers to another GIIS) to construct a decentralized hierarchy of information. It is possible to distribute a Registry, but there is no standard protocol to distribute multiple Registries. Hawkeye does not support a decentralized architecture (e.g., there is no mechanism to link different pools, although it is possible in future development), which may make it more vulnerable to failure.

How the data is communicated, using a push or a pull mechanism also differs. The MDS allows only a pull model for data transfer. This can cause a bottleneck between client and server, and may not meet the needs of a user who would rather have a notification or push model. R-GMA supports both the pull and the push models for transferring the information data. For Hawkeye, executing a job based on a resource attribute value is similar to a Producer initiating contact with a Consumer. The administrator (the interested consumer) submits a "trigger" ClassAd to the Manager and when the requested data is available (i.e., the event specified in the "trigger" ClassAd occurs) the administrator is notified by email (or any other requested job).

## 4. Conclusion and Future Work

We have investigated three monitoring and information services for distributed systems: MDS, R-GMA and Hawkeye. A collection of experiments was performed on each service to study their performance behaviors and constraints. Our work shows that all three services present good scalability in most of the cases we evaluated. In the four sets of experiments we conducted to evaluate the performance of the service components under different circumstances, we found a strong advantage to caching or pre-fetching the data, as well as the need to have primary components at well connected sites due to high load seen by all systems.

In our future work, we plan to do more experiments to study other characteristics of these services. For example, the testbeds in our study were built in a LAN environment; the experiments should be repeated to study performance in a WAN environment. We also need to determine the difference between querying an aggregate information server and an information server for the same piece of information. We plan to consider additional patterns of user access.

## ACKNOWLEDGMENTS

We thank John Mcgee, ISI, for assistance with the MDS; James Magowan, IBM-UK, for assistance with R-GMA; and both Alain Roy and Nick Leroy at the University of Wisconsin, Madison, for assistance with Hawkeye. We also thank Scott Gose and Charles Bacon for assistance with the testbed at Argonne. This work was supported in part by the Mathematical, Information, and Computational Sciences Division subprogram of the Office of Advanced Scientific Computing Research, U.S. Department of Energy, under contract W-31-109-Eng-38.

## REFERENCES

[1] Aloisio, G., M. Cafaro, I. Epicoco, and S. Fiore, "Analysis of the Globus Toolkit Grid Information Service". Technical report GridLab-10-D.1-0001-GIS_Analysis, GridLab project, http://www.gridlab.org/Resources/Deliverables/D10.1.pdf.

[2] Czajkowski, K., S. Fitzgerald, I. Foster, and C. Kesselman, "Grid Information Services for Distributed Resource Sharing". In Proc. 10$^{th}$ IEEE International Sym.p. on High Performance Distributed Computing (HPDC-10), IEEE Press, 2001.

[3] DataGrid: http://eu-datagrid.web.cern.ch/eu-datagrid/

[4] "DataGrid Information and Monitoring Services Architecture: Design, Requirements and Evaluation Criteria", Technical Report, DataGrid, 2002.

[5] Fisher, S., "Relational Model for Information and Monitoring". Technical Report GWD-Perf-7-1, GGF, 2001.

[6] Foster, I., and C. Kesselman, "Globus: A Metacomputing Infrastructure Toolkit".Intl J. Supercomputer Applications, 11(2):115-128, 1997.

[7] Foster, I., and C. Kesselman, eds, "The Grid: Blueprint of a New Computing Infrastructure", Morgan Kaufmann, 1999. ISBN 1-55860-475-8.

[8] Ganglia: http://ganglia.sourceforge.net

[9] Global Grid Forum: http://www.gridforum.org/

[10] Hawkeye: http://www.cs.wisc.edu/condor/hawkeye

[11] Java™ Servlet Technology: http://java.sun.com/products/servlet

[12] Knop, M., J. Schopf, and P. Dinda, "Windows Performance Monitoring and Data Reduction using WatchTower", 11th IEEE Symposium on High-Performance Distributed Computing (HPDC11), 2002.

[13] Litzkow M., M. Livny, and M. Mutka, "Condor - A Hunter of Idle Workstations". In Proceedings of the 8th International Conference of Distributed Computing Systems, pages 104-111, June 1988.

[14] MDS: http://www.globus.org/mds/

[15] OpenLdap: http://www.openldap.org/

[16] Plale, B., "Whitepaper on Synthetic Workload for Grid Information Services/Registries". DataWorkshop 2003 held in conjunction with GlobusWorld 2003, San Diego, 2003.

[17] Plale, B., P. Dinda, and G. Laszewski, "Key Concepts and Services of a Grid Information Service". ISCA 15th International Parallel and Distributed Computing Systems (PDCS), 2002.

[18] Raman, R., "Matchmaking Frameworks for Distributed Resource Management". PhD thesis, Department of Computer Science, University of Wisconsin, 2001.

[19] Raman R., M. Livny, and M. Solomon, "Matchmaking: Distributed resource management for high throughput computing". In Proceedings of the Seventh IEEE International Symposium on High Performance Distributed Computing (HPDC7), July 1998.

[20] Tierney, B., R. Aydt, D. Gunter, W. Smith, V. Taylor, R. Wolski, and M. Swany, "A Grid Monitoring Architecture". Global Grid Forum Performance Working Group, 2001.

[21] Yeong, W., T. Howes, and S. Kille, "Lightweight Directory Access Protocol", RFC 1487. In Performance Systems International, University of Michigan, ISODE Consortium, July 1993.